# Short-term Forecasting of Anomalous Load Using Rule-based Triple Seasonal Methods

Siddharth Arora, James W. Taylor.
*IEEE Transactions on Power Systems*, forthcoming.

*Abstract*—Numerous methods have been proposed for forecasting load for normal days. Modeling of anomalous load, however, has often been ignored in the research literature. Occurring on special days, such as public holidays, anomalous load conditions pose considerable modeling challenges due to their infrequent occurrence and significant deviation from normal load. To overcome these limitations, we adopt a rule-based approach, which allows incorporation of prior expert knowledge of load profiles into the statistical model. We use triple seasonal Holt-Winters-Taylor (HWT) exponential smoothing, triple seasonal autoregressive moving average (ARMA), artificial neural networks (ANNs), and triple seasonal intraweek singular value decomposition (SVD) based exponential smoothing. These methods have been shown to be competitive for modeling load for normal days. The methodological contribution of this paper is to demonstrate how these methods can be adapted to model load for special days, when used in conjunction with a rule-based approach. The proposed rule-based method is able to model normal and anomalous load in a unified framework. Using nine years of half-hourly load for Great Britain, we evaluate point forecasts, for lead times from one half-hour up to a day ahead. A combination of two rule-based methods generated the most accurate forecasts.

*Index Terms*—Anomalous load, forecasting, rule-based approach, triple seasonality.

## I. INTRODUCTION

SHORT-TERM forecasts of electricity demand (*load*) are crucial for real-time scheduling of power systems, optimizing operational costs, and improving the reliability of distribution networks. A slight increase in the load forecast error leads to a considerable increase in operating costs for an electric utility [1]. Due to the impact of load forecasts on the reliability of power systems and its financial implications for energy markets, it is imperative to model load accurately.

Numerous statistical methods have been proposed for short-term load forecasting [2], [3], whereby the lead times under consideration vary from a few minutes up to a day ahead. The presence of intraday and intraweek seasonal cycles is a prominent feature of load, and statistical time series methods aim to capture this 'double seasonality'. Such methods have been shown to perform well for normal days. However, they are unlikely to be of use for special days with anomalous load, such as public holidays and long weekends. The load profile (shape of intraday load curve) differs for different types of special day. Also, for a given special day, such as a particular public holiday, the profile may vary from one year to the next, based on the day of week and the time of the year on which the special day occurs. Due to these complexities, forecasting load under anomalous conditions has largely been omitted from time series forecasting methods [2], and has hence been left to the experience and judgment of the central controller of the electricity grid [4]-[6]. The load observed on normal working days is referred to as *normal load*, whereas the load observed on special days is referred to as *anomalous load*.

Previous approaches for short-term forecasting of normal and anomalous load have mostly employed regression-based methods, whereby special day effects are incorporated using dummy variables [7]-[9]. Some authors have proposed rule-based approaches [4]-[6], while others have used ANNs [10]-[14]. To avoid the model becoming over-parameterized, different special days have often been classified as belonging to the same special day type [7]-[9]. This relies on the assumption that the load profile for different special days can be treated as similar, and would remain similar over the years. Given that the pattern of load for each special day tends to change each year, this assumption is too restrictive.

We propose a rule-based approach for load forecasting. It has been shown that, when domain knowledge is available and the time series has a consistent structure (seasonality in this case), rule-based forecasting can outperform conventional extrapolation methods [15], [16]. In our work, the rule essentially identifies the past special day that has daily load profile that will be most useful in enhancing the time series model's estimate of load for the future special day. The formulation of rules is done subjectively by inferring them directly from the data. Once developed, the rules are flexible enough to incorporate additional knowledge arising from the inclusion of new observations.

For exponential smoothing, ARMA, and ANN methods, it has been shown recently that accommodating the intrayear seasonality in load, along with the intraday and intraweek seasonal cycles, leads to improved short-term load forecast accuracy [17]. This triple seasonal framework is convenient for our modeling of anomalous load, because special days tend to reoccur annually. Note that the existing methods have been applied to load time series for which special days are either treated as a Sunday [2], or smoothed out [3]. Our initial focus in this paper is a new rule-based HWT exponential smoothing approach, which models normal and anomalous load in a unified framework. After introducing this approach, we consider analogous rule-based ARMA, ANN and SVD-based exponential smoothing methods.

Section II describes the load characteristics. Section III presents the rule-based adaptations of triple seasonal HWT exponential smoothing, ARMA, ANN and SVD-based exponential smoothing. Section IV formulates the rules. Empirical results are provided in section V. Section VI summarizes and concludes the paper.

## II. LOAD CHARACTERISTICS

We used nine years of half-hourly load for Great Britain from 1st Jan. 2001 to 31st Dec. 2009. The first eight years of the data were used for estimating the model parameters, and observations from the final year were used for model evaluation. We generated point forecasts using a rolling forecast origin though the post-sample data, for horizons varying from one half-hour up to a day ahead.

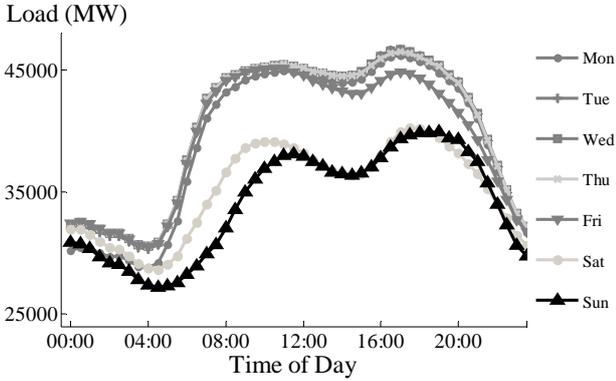

Fig. 1. Average intraday profile for each day of the week.

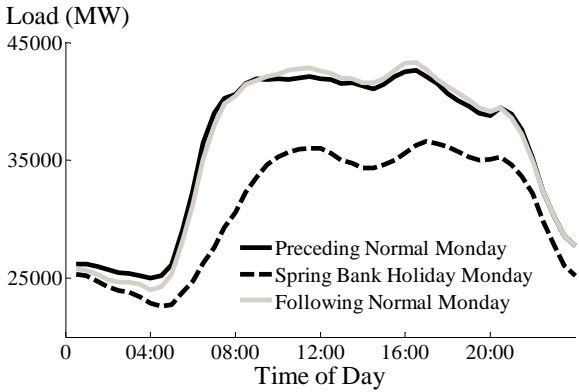

Fig. 2. Load profile for a Spring Bank Holiday Monday (26 May 2008) and a normal Monday from the preceding and following week.

Fig. 1 presents the average intraday load cycle for each day of the week, calculated using the estimation sample. Load is clearly much lower at the weekends than on the weekdays. The load profile for Saturdays differs from Sundays, and for the weekdays, there are some differences, most notably on Mondays and Friday afternoons. Along with daily (intraday) and weekly (intraweek) cycles, load also exhibits large annual variations (intrayear cycles) due to weather and seasonal variations across the year. The methods used in this work are aimed at capturing the 'triple seasonality' in load.

To model load adequately, special day effects need to be taken into account, along with seasonality. Fig. 2 plots load on a special day, namely a bank holiday Monday (a public holiday in Great Britain), and a normal Monday from the preceding and following weeks. It can be seen that load on the special day is noticeably lower compared to the normal days.

Prior to modeling, we applied the natural log transformation to the data, in order to convert multiplicative effects to additive. We represent the length of the intraday, intraweek and intrayear seasonal cycles by $m_1$, $m_2$ and $m_3(t)$ respectively. We have $m_1$=48, and $m_2$=336. Following [17], we set the length of the intrayear seasonal cycle, for period $t$, to be $m_3(t)$=52×336, except for some years, where we set $m_3(t)$=53×336 for a few weeks around the clock-change.

## III. FORECASTING METHODS

### A. Triple Seasonal HWT Exponential Smoothing

Triple seasonal HWT exponential smoothing has recently been proposed for modeling the intraday, intraweek and intrayear seasonality in intraday load data [17]. The method is represented in state space form as follows:

$$y_t = l_{t-1} + d_{t-m_1} + w_{t-m_2} + a_{t-m_3(t)} + \phi e_{t-1} + \varepsilon_t \quad (1)$$
$$e_t = y_t - (l_{t-1} + d_{t-m_1} + w_{t-m_2} + a_{t-m_3(t)}) \quad (2)$$
$$l_t = l_{t-1} + \lambda e_t \quad (3)$$
$$d_t = d_{t-m_1} + \delta e_t \quad (4)$$
$$w_t = w_{t-m_2} + \omega e_t \quad (5)$$
$$a_t = a_{t-m_3(t)} + \alpha e_t \quad (6)$$

where $\varepsilon_t \sim NID(0, \sigma^2)$, and $\sigma^2$ denotes a constant variance. The log transformed load is denoted by $y_t$; $l_t$ is the smoothed level; the intraday seasonal index is denoted by $d_t$; and $w_t$ is the seasonal index for the intraweek cycle that remains after the intraday seasonality has been removed. The seasonal index $a_t$ denotes the intrayear cycle, which remains after both the intraday and the intraweek seasonality have been removed. The term $m_3(t)$ is the intrayear cycle length, for period $t$. The smoothing parameters are denoted by $\lambda$, $\delta$, $\omega$, and $\alpha$. The term involving the parameter $\phi$ adjusts for autocorrelation in the error $e_t$. Initial seeds for the level and seasonal indices were calculated using the few years of data, as done in [17]. Model parameters were estimated by maximizing a Gaussian likelihood, which for this model is equivalent to minimizing the sum of squared one-step ahead errors. The model is suitable only for load with special days smoothed out.

To model normal and anomalous load in a unified framework, we present a rule-based modified HWT (RB-HWT) exponential smoothing formulation as follows:

$$y_t = l_{t-1} + d_{t-m_1} + w_{t-m_2} + a_{t-m_3(t)} + \phi e_{t-1}$$
$$\quad + I_{N_t} \varepsilon_t^{(N)} + I_{S_t} \varepsilon_t^{(S)} \quad (7)$$
$$e_t = y_t - (l_{t-1} + d_{t-m_1} + w_{t-m_2} + a_{t-m_3(t)}) \quad (8)$$
$$l_t = l_{t-1} + \lambda e_t \quad (9)$$
$$d_t = d_{t-m_1} + I_{N_t} \delta e_t \quad (10)$$
$$w_t = w_{t-m_2} + I_{N_t} \omega e_t \quad (11)$$
$$a_t = a_{t-m_3(t)} + I_{S_t} \alpha_1 e_t + I_{S_t} \alpha_2 e_t \quad (12)$$

where $\varepsilon_t^{(N)} \sim NID(0, \sigma_N^2)$ and $\varepsilon_t^{(S)} \sim NID(0, \sigma_S^2)$ are the model errors for normal and special days, respectively, having corresponding variances $\sigma_N^2$ and $\sigma_S^2$. The binary indicator term $I_{N_t}$ equals one if $t$ occurs on a normal day, and zero otherwise, whereas $I_{S_t}$ equals one for special days, and zero otherwise. Thus, at any given period $t$, we have $I_{S_t} = 1 - I_{N_t}$.

Fig. 3 presents the intraday, intraweek and intrayear seasonal indices. As shown in Fig. 3a, the intraday and intraweek seasonal indices, denoted by $d_t$ and $w_t$, capture the daily and weekly patterns in load for normal days. In Fig. 3b, the index $a_t$ denotes the intrayear cycle, which remains after both the intraday and intraweek seasonality for only the normal days has been removed. Thus, $a_t$ incorporates the intrayear seasonal pattern in load for all days, along with the intraday and intraweek seasonal load patterns for special days. Specifically, the difference between a normal and anomalous load profile for the same day of the week is embedded and modeled via $a_t$.

In Fig. 3b, the anomalous load periods are denoted in black. It is noteworthy that $a_t$ on a special day is considerably lower compared to a normal day from the adjacent weeks. Inspection of the data reveals that load for the special days is noticeably lower than normal load for the same day of the week, around the same date. In Great Britain, the special days are New Year's day, the day following New Year's day, Good Friday, Easter Monday, three summer bank holiday Mondays, and the Christmas period, which stretches from 21 December to 31 December, inclusive. We refer to Good Friday and Easter Monday as Easter holidays.

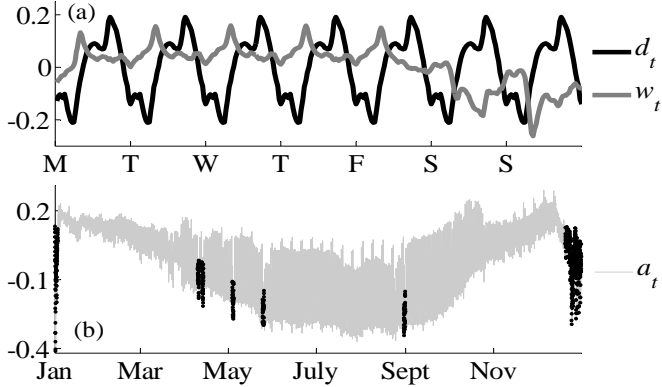

Fig. 3. Seasonal indices for the intraday, intraweek, and intrayear cycle, denoted by $d_t$, $w_t$ and $a_t$, respectively. In panel (b), the grey line denotes the index on normal days, while black dots correspond to anomalous periods.

We update $a_t$ at different rates for the two day types. This allows the impact of the annual seasonal cycle on normal and anomalous load to be different. The indices $d_t$ and $w_t$ are not updated on special days, as they capture the seasonal patterns for only the normal days. Crucially, the intrayear cycle length, denoted by $m_3(t)$, is determined using a rule-based approach for special days, which is discussed in Section IV. This cycle length allows the selection of suitable historical anomalous load observations in the statistical model, to be extrapolated into the future for generating forecasts. For normal days, $m_3(t)$ is chosen to be the same as the intrayear cycle length used in the original triple seasonal HWT method.

For periods on normal days, the RB-HWT model defaults to the original HWT model. For special days, the RB-HWT model incorporates the daily and weekly seasonal effects in normal load, and adjusts it using the intrayear seasonal index, which captures the variation in anomalous load from normal load, for each special day. This allows the model to treat each special day as having its own unique load profile.

The maximum likelihood estimation uses a Gaussian error, for which the variance is different on special days to that on normal days. The log-likelihood ($LL$) can be written as:

$$LL = -\frac{n_{ND}}{2}\log(2\pi\sigma_N^2) - \frac{n_{SD}}{2}\log(2\pi\sigma_S^2) - \sum_{t=365\times m_1+1}^{N}\left(\frac{I_{N_t}}{2\sigma_N^2}(\varepsilon_t^{(N)})^2 + \frac{I_{S_t}}{2\sigma_S^2}(\varepsilon_t^{(S)})^2\right) \tag{13}$$

where $N$ is the size of the estimation sample, and $n_{ND}$ and $n_{SD}$ are the number of observations that fall on normal and special days, respectively, in the estimation sample, excluding observations from the first year.

### B. Triple Seasonal ARMA

In [17], the following triple seasonal ARMA model has been proposed for modeling load on normal days,

$$Y_p(L)\Phi_{P_1}(L^{m_1})X_{P_2}(L^{m_2})\Psi(L^{m_3(t)})(y_t - c) = \Omega_q(L)\Theta_{Q_1}(L^{m_1})\Gamma_{Q_2}(L^{m_2})\Lambda(L^{m_3(t)})\varepsilon_t \tag{14}$$

where $c$ is a constant; $\varepsilon_t$ is the model error; $L$ is the lag operator, and $Y_p$, $\Phi_{P_1}$, $X_{P_2}$, $\Omega_q$, $\Theta_{Q_1}$ and $\Gamma_{Q_2}$ are polynomial functions of order $p$, $P_1$, $P_2$, $q$, $Q_1$, and $Q_2$. We considered orders equal to or less than three. The functions $\Psi$ and $\Lambda$ essentially capture the annual cycles. $\Psi$ is written as:

$$\Psi(L^{m_3(t)}) = 1 + \eta_1 L^{m_3(t)} + \eta_2 L^{m_3(t) + m_3(t - m_3(t))} + \eta_3 L^{m_3(t) + m_3(t - m_3(t)) + m_3(t - m_3(t - m_3(t)))} \tag{15}$$

where $\eta_1$, $\eta_2$ and $\eta_3$ are constant coefficients. The function $\Lambda$ includes the same lags as $\Psi$, but comprises different coefficients. To extend the model for anomalous load, as we did for the HWT model, we used a rule-based approach to set the length of the lag $m_3(t)$ for the intrayear cycle for special days. We refer to the new model as rule-based SARMA (RB-SARMA), and represent it as:

$$Y_p(L)\Phi_{P_1}(L^{m_1})X_{P_2}(L^{m_2})\left(I_{N_t}\Psi(L^{m_3(t)}) + I_{S_t}\theta(L^{m_3(t)})\right)(y_t - c) = \Omega_q(L)\Theta_{Q_1}(L^{m_1})\Gamma_{Q_2}(L^{m_2})\left(I_{N_t}\Lambda(L^{m_3(t)}) + I_{S_t}K(L^{m_3(t)})\right)\left(I_{N_t}\varepsilon_t^{(N)} + I_{S_t}\varepsilon_t^{(S)}\right) \tag{16}$$

where $\varepsilon_t^{(N)} \sim NID(0, \sigma_N^2)$ and $\varepsilon_t^{(S)} \sim NID(0, \sigma_S^2)$ are the model errors for normal and special days, respectively, while $\theta$ and $K$ are additional functions of the lag operator $L^{m_3(t)}$ used only for special days. Hence, we basically switch between different AR and MA polynomial functions, with different annual lag terms, depending on whether $t$ belongs to a normal day, or a special day. For special days, the rule-based value for $m_3(t)$ allows the inclusion of load observations from three previous special days, which would be suitable for improving the model's forecast for a given special day. As with the RB-HWT method, the parameters of the RB-SARMA method are estimated using a maximum likelihood procedure as used for the RB-HWT method. The Box-Jenkins methodology was used to select polynomial function orders for AR and MA terms.

### C. ANN

We include ANNs in this study, as they have previously been employed for modeling anomalous load [10]-[14]. We extend the triple seasonal ANN method, as presented in [17], using a rule-based approach to model load for normal and special days. To choose a suitable ANN for this data, we closely followed the ANN architecture used by [17] for forecasting load for Great Britain. Specifically, we employ a univariate feed-forward ANN with a single hidden layer and a single output. Note that univariate models use only historical load observations, and have been shown to be competitive with weather-based models for short-term load forecasting [3]. As a pre-processing step, we used a differencing operator of the form $(1-L^{m_1})(1-L^{m_2})$, as this led to an improvement in the forecast accuracy over the cross-validation hold-out sample, which we set as the final year of the estimation sample.

Since ANNs have been shown to struggle with multi-step-ahead prediction [18], we build a separate ANN model for each horizon. We specify the output to comprise load observations, differenced using the above operator, and normalized to have zero mean and unit standard deviation. We selected the lags for the input variables to be as consistent as possible with the SARMA model. Specifically, for the ANN built for horizon $h$, we used load at the forecast origin, and at the following lags: 1, 2, $m_1$-$h$, $2m_1$-$h$, $3m_1$-$h$, $m_2$-$h$, $2m_2$-$h$, $3m_2$-$h$, $LagA=m_3(t)$-$h$, $LagB=LagA+m_3(t-LagA-h)$, and $LagC = LagB+m_3(t-LagB-h)$. Using this model, $m_3(t)$ is defined in the same way as in the exponential smoothing and SARMA models of the previous two subsections. We refer to this as the rule-based ANN, denoted by RB-ANN.

We used least squares with backpropagation to estimate the link weights. The activation functions for the hidden and output layer were chosen to be sigmoid and linear, respectively. We selected the input variables, number of units in the hidden layer, backpropagation learning rate and momentum parameters, and regularization parameters, using only the cross-validation hold-out data.

### D. Triple Seasonal SVD-based Exponential Smoothing

SVD-based approaches have been proposed for forecasting intraday time series, see [3], [19]. The rationale of SVD-based methods lies in transforming the data to an orthogonal space, and allowing the focus to be on forecasting only the components that capture a major proportion of variance in the data, thereby reducing the dimensionality of the model.

In [3], SVD is applied to the data arranged in the form of a matrix $Y_{w \times m_2}$, where $w$ denotes the number of weeks in the estimation data. Applying SVD to $Y$ gives, $Y = USV'$, where $U_{w \times w}$ and $V_{m_2 \times m_2}$ are orthogonal matrices. $S_{w \times m_2}$ is a diagonal matrix constituting singular values. The data matrix is projected onto $V$, using $P = YV$. The columns of $V$ and $P$ are termed the *intraweek feature vectors* and *interweek feature series*, respectively. The matrix $Y$ can be reconstructed using $Y = PV'$. Dimension reduction is achieved by using only $k$ ($k < m_2$) feature vectors and series in the modeling.

For modeling anomalous load, we extend the method used in [3], and propose a rule-based modified intraweek SVD (RB-SVD) exponential smoothing method as follows:

$$y_t = \tilde{p}_{t-1}\tilde{V}'_{[t \bmod m_2]} + a_{t-m_3(t)} + \phi e_{t-1} + I_{N_t}\varepsilon_t^{(N)} + I_{S_t}\varepsilon_t^{(S)} \tag{17}$$

$$e_t = y_t - (\tilde{p}_{t-1}\tilde{V}'_{[t \bmod m_2]} + a_{t-m_3(t)}) \tag{18}$$

$$\tilde{p}_t = \tilde{p}_{t-1} + (\alpha 1_{m_2}\tilde{V} + I_{N_t}\delta\sum_{j=1}^{7}\tilde{V}_{[t \bmod m_1]+(j-1)m_1} + I_{N_t}\omega\tilde{V}_{[t \bmod m_1]})e_t \tag{19}$$

$$a_t = a_{t-m_3(t)} + I_{N_t}\alpha_1 e_t + I_{S_t}\alpha_2 e_t \tag{20}$$

where $\varepsilon_t^{(N)} \sim NID(0, \sigma_N^2)$ and $\varepsilon_t^{(S)} \sim NID(0, \sigma_S^2)$ are the model errors for normal and special days, respectively, while $I_{N_t}$ and $I_{S_t}$ are dummy variables, as defined earlier for RB-HWT. $\tilde{p}_t$ is a row vector of length $k$ comprising values in the first $k$ interweek feature series in period $t$, $\tilde{V}_{m_2 \times k}$ comprises the first $k$ intraweek feature vectors, $\tilde{v}_{[s]}$ is a row of $\tilde{V}$ having index $s$, and $1_{m_2}$ is a row vector of ones having length $m_2$. The term $a_t$ accommodates the special day effects using a rule-based $m_3(t)$, as explained in the previous sub-section. The model parameters were estimated using the maximum likelihood procedure as used for the RB-HWT method. We estimated $k$ using cross-validation.

The parameters in equation (19) are similar to the smoothing parameters used in the HWT framework. The parameter $\lambda$ updates the level of $\tilde{p}_t$, $\delta$ takes into account the intraday seasonal effects, while $\omega$ accommodates the intraweek seasonal effects, and updates $\tilde{p}_t$ based on the period of the week on which $t$ occurs. The index $a_t$ accommodates the special day effects using a rule-based $m_3(t)$, as explained in the previous sub-section. The model parameters were estimated using the maximum likelihood procedure as used for the RB-HWT method. We estimated $k$ using cross-validation.

## IV. FORMULATING RULES

A rule-based approach allows the encapsulation of domain knowledge and data characteristics into a statistical modeling framework [15],[16]. The prior knowledge and data properties are quantified and expressed explicitly in the form of rules. In this paper, as stated in Section III, the rules have the sole purpose of defining the most suitable length of the annual lag $m_3(t)$ for special days, in the formulations of the HWT, SARMA, ANN and SVD models. In this section, we present four alternative rules for defining $m_3(t)$ for special days.

We present load for different special days in Fig. 4. The figure shows that different special days should not be treated as having similar load profiles. Thus, we treat each special day separately, whereby each special day is allowed to have a unique load profile, which may change over the years.

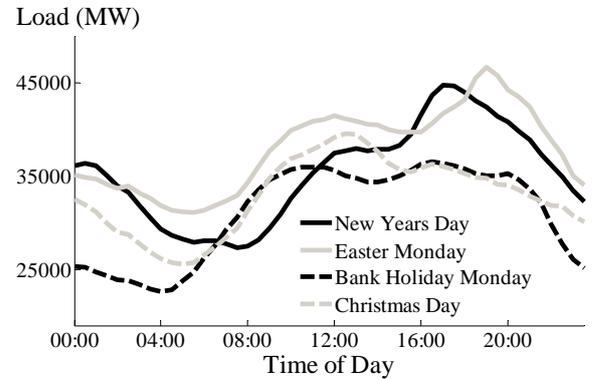

Fig. 4. Load profile for New Year's Day (1 January), Easter Monday (24 March), Spring Bank Holiday Monday (26 May), and Christmas Day (25 December) observed in the year 2008.

The special days are of two types. The first type includes those special days that fall on the same day of the week each year, e.g. Good Friday. The second type includes those days that occur at the same date each year, e.g. Christmas Day. We formulate different rules for the two special day types, using only the estimation data. The calculation of intrayear cycle length $m_3(t)$ takes into account the inclusion of extra observations due to leap years, for all rules proposed in this study. Note that for a given rule to work, the length of the time series should either be equal or greater than the largest $m_3(t)$ selected using that rule.

### A. Rule 1

To model load for a given special day, this rule refers to the historical load observed on the same special day from the previous year. For example, to forecast load for Good Friday in 2008 (21 March 2008), $m_3(t)$ is selected such that observations from the previous year's Good Friday (6 April 2007) are included in the model. This leads to $m_3(t) = 350 \times 48$ for each period $t$ on the Good Friday in 2008. Similarly, to model load for Christmas Day, $m_3(t)$ refers to Christmas Day in the previous year. Thus, $m_3(t)$ is defined such that $y_t$ and $y_{t-m_3(t)}$ belong to: (a) the same special day, (b) adjacent years, and, (c) the same period of the day.

A potential disadvantage of Rule 1 is that it ignores the day of the week on which the special day occurs. For example, Christmas Day in the current year obviously falls on a different day to Christmas Day in the previous year (e.g. Thursday in 2008 and Tuesday in 2007). Given that the average intraday cycle is not the same for each day of the week (see Fig. 1), it is worth investigating if different days of the week have different impact on anomalous load. Also, it has been pointed out that if a particular special day falls, in different years, on the same day of the week, then the load pattern will be very similar [12]. Thus, in the next rule, we aim to accommodate the intraweek seasonal effects on anomalous load.

### B. Rule 2

This rule treats each day of the week as having a unique impact on anomalous load. Rule-based approaches for incorporating the day of the week effect on anomalous load have been employed by [5], [6]. Using this rule, $m_3(t)$ is selected such that $y_t$ and $y_{t-m_3(t)}$ belong to the same: (a) special day, (b) day of the week, and, (c) period of the day. For example, to model load for a New Year's day, $m_3(t)$ refers to a previous year's New Year's Day which occurred on the same day of the week as the current day. For example, consider New Year's Day in 2008. This was a Tuesday, and so $m_3(t)$ is defined to be such that we are referring back to the most recent occurrence of New Year's Day on a Tuesday, which was in 2002. Therefore, for New Year's Day in 2008, we set $m_3(t) = (5 \times 365 + 366) \times 48$. For special days that fall on the same day of the week each year, namely the three summer Bank holiday Mondays and Easter holidays, this rule is the same as Rule 1.

A disadvantage of Rule 2 is that it can lead to the selection of a large value for $m_3(t)$, which implies long lags and hence the inclusion of less recent observations in the model. It has been argued that economic factors and weather conditions can alter the load patterns over time [5], [6], [12], which motivates the use of more recent observations in forecasting. Also, in some cases, the magnitude of $m_3(t)$ tends to be larger than the total number of historical observations required to use this rule. In such cases, Rule 2 reverts to Rule 1.

In situations where an insufficient number of observations are available, a special day has been treated as another special day type by [5], [6], while different special days have often been classified as being the same [7]-[9]. Hence, to reduce the magnitude of $m_3(t)$, for Rule 2, we classify special days within the Christmas period, as having the same profile. This also allows us to investigate the impact, on forecast accuracy, of treating different special days, as being the same. The special days that do not belong to the Christmas period, namely the New Year's day, day following New Year's day, Easter holidays, and bank holidays Mondays, are all treated separately, whereby each special day is allowed to have a unique profile.

### C. Rule 3

Given that weekdays witness a relatively high load compared to weekends (Fig. 1), this rule treats weekdays as having a distinct impact on anomalous load, in comparison with weekends. Specifically, for special days that fall on the same date each year, $m_3(t)$ is selected such that $y_t$ and $y_{t-m_3(t)}$: (a) belong to the same special day, (b) either both fall on a weekday or both fall on a weekend, and, (c) belong to the same period of the day. For example, to forecast load for Christmas Day in 2009, which fell on a Friday, we set $m_3(t) = 365 \times 48$, because the previous year's Christmas Day also fell on a weekday. As another example, consider Boxing Day in 2009. This was a Saturday, and so $m_3(t)$ must be defined so that we are referring back to a previous Boxing Day that fell on a weekend, and this was Boxing Day in 2004, which fell on a Sunday. Therefore, for the periods on Boxing Day 2009, we set $m_3(t) = (4 \times 365 + 366) \times 48$.

For the three summer bank holiday Mondays, this rule is the same as Rule 1. For the Easter holidays, $m_3(t)$ is selected using Rule 1, with one extra condition, that both $y_t$ and $y_{t-m_3(t)}$ occur either before, or after, the summertime clock-change in that year. For example, to forecast load for Good Friday in 2008 (21 March 2008), which occurs before the summertime clock-change, $m_3(t)$ refers to Good Friday in 2005 (25 March 2005), because this was the most recent Good Friday that had also occurred before the summertime clock-change. The summertime clock-changes in 2005 and 2008 occurred at 27 March and 30 March, respectively. This extra condition on Rule 1 was motivated by us observing that the daily load can change quite noticeably at clock-change.

We refer to the special days from 21 to 24 December, and from 27 to 30 December, inclusive, as proximity days. We use the notation PD for these days. They tend to witness lower load than normal days, but higher than other special days within the Christmas period. We define $m_3(t)$ differently for two separate classes of PDs, based on whether they precede a Christmas Day, or follow a Boxing Day. Furthermore, a PD is classified as a bridging proximity day, with the notation B-PD, if it is the only day occurring between a weekend and either a Christmas Day or a Boxing Day. Otherwise, it is treated as a non-bridging proximity day, *NB-PD*. Overall, the load on a *PD* tends to be lowered further if it is a *B-PD*, and hence we treat them as being separate from *NB-PD*, as also done by [8]. For a *PD*, Rule 3 sets $m_3(t)$ such that both $y_t$ and $y_{t-m_3(t)}$: (a) precede a Christmas Day or follow a Boxing day, (b) occur either on a weekend or a weekday, (c) either belong to the class *B-PD* or *NB-PD*, and, (d) belong to the same period of the day. For example, for Saturday 29 December 2007, we set $m_3(t) = 364 \times 48$, so that the lag refers to Saturday 30 December 2006. This is because both: (a) followed a Boxing Day, (b) occurred at a weekend, and, (c) belonged to the class *NB-PD*. As another example, consider Monday 24 December 2007. For this proximity day, we set $m_3(t) = (5 \times 365 + 366) \times 48$ so that the lag refers to Monday 24 December 2001. This is because both: (a) preceded a Christmas Day, (b) occurred on a weekday, and, (c) belonged to the class *B-PD*. In cases where there are no historical load observations that satisfy this rule, we revert to Rule 1.

This rule incorporates recent information in the model by utilizing the similarity in the intraday cycle, models proximity days separately, and incorporates the effect of bridging days on anomalous load. However, this rule treats all weekdays as having the same intraday cycle. It can be seen from Fig. 1, that not all weekdays have the same average intraday cycle. Also, the average intraday cycle for a Saturday is different from a Sunday. To overcome these limitations, we formulate Rule 4.

### D. Rule 4

This rule treats different days of a week, which have a distinct intraday cycle, as having a distinct impact on anomalous load. A week is treated as comprising five different intraday cycles, as used in [3]. Specifically, Tuesday, Wednesday and Thursday are treated as having a common intraday cycle, whereas Monday, Friday, Saturday and Sunday are each assumed to have a distinct intraday cycle.

This rule is similar to Rule 3, with the key difference that, for all special days that occur at the same date each year, instead of selecting an historical special day based on whether it occurs on a weekday or a weekend, this rule selects an historical special day of the same class of intraday cycle as the current special day. For example, consider Christmas Day in 2007. This was a Tuesday, and so $m_3(t)$ is defined to be such that we are referring to the most recent occurrence of Christmas Day with the same intraday cycle type as Tuesday, and this was Christmas Day in 2003, which fell on a Thursday. This is because Tuesday and Thursday are treated as having the same intraday cycle using this rule. Hence, for periods on Christmas Day 2007, we set $m_3(t) = (3 \times 365 + 366) \times 48$. For special days that fall on the same day of the week each year, such as Good Fridays, this rule is the same as Rule 3.

### V. EMPIRICAL COMPARISON OF METHODS

The main focus of our empirical analysis was the special days. In all, there were eighteen special days in the one-year post-sample period. We generated point forecasts for all days in the post-sample period, and evaluated accuracy using the mean absolute percentage error (MAPE) and root mean squared percentage error (RMSPE). The relative model rankings were similar for the two measures; hence we present results only for MAPE in this paper, which is computed as:

$$MAPE_h = \frac{1}{N-T-h+1} \sum_{i=T+h}^{N} \left| \frac{\exp(y_i) - \exp(\hat{y}_i)}{\exp(y_i)} \right| \qquad (21)$$

where $MAPE_h$ is the MAPE at horizon $h$, $\exp(y_i)$ is the actual load, $\exp(\hat{y}_i)$ is the corresponding forecast, $T$ is the forecast origin, and $N$ is the length of the time series. We present the forecast accuracy of the four univariate methods presented in Section III, when used in conjunction with the four different rules described in Section IV. As benchmarks, we include the four univariate methods without any adjustment for special days. This amounts to treating the special days as normal days, with no domain knowledge incorporated in the models.

Fig. 5 presents the HWT results for just the special days. The figure shows the original HWT method comfortably outperformed by all the RB-HWT approaches, apart from the one based on Rule 2, which performed poorly. The best performance was produced by Rule 3. This set of results highlights the value in using a subjective rule-based approach in situations where domain knowledge is available.

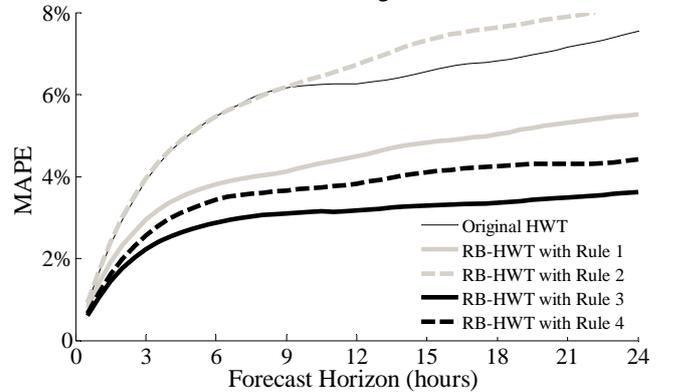

Fig. 5. MAPE across special days for the original and rule-based HWT.

In Fig. 6, we present the SARMA results for just the special days. The figure shows that, as with the RB-HWT methods, Rule 3 delivered the best results, although its superiority over Rules 1 and 4 was small. As with RB-HWT, Rule 2 was disappointing for the RB-SARMA method. A possible reason for this is that, unlike the other rules, Rule 2 treats different special days within the Christmas period as being the same.

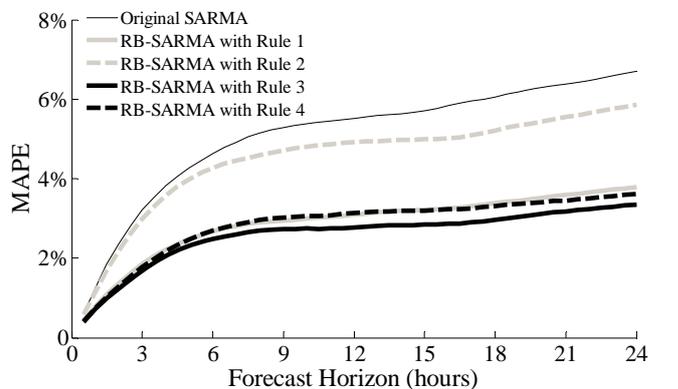

Fig. 6. MAPE across special days for the original and rule-based SARMA.

Fig. 7 shows the ANN results for just the special days. The most accurate results correspond to RB-ANN with Rules 1 and 3. The original ANN method was less accurate than all the RB-ANN methods, except when Rule 2 and 4 were used.

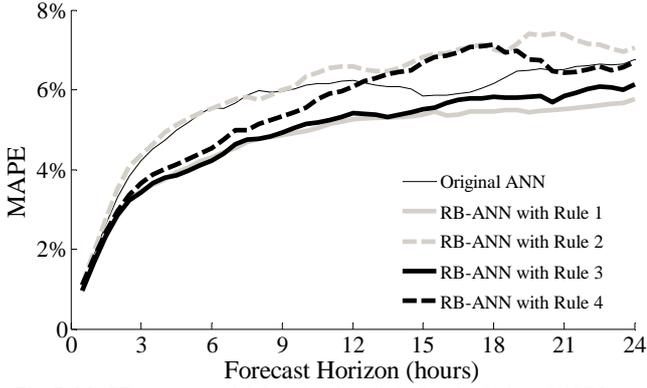

Fig. 7. MAPE across special days for the original and rule-based ANN.

In Fig. 8, we present the SVD results for just the special days. The relative rankings of the RB-SVD methods is very similar to that of RB-HWT, whereby Rule 3 provides the best results, while Rule 2 gives the worst performance. For the best performing SVD method, i.e. RB-SVD with Rule 3, cross-validation indicated $k = 29$ feature vectors and series, which implies quite substantial dimension reduction.

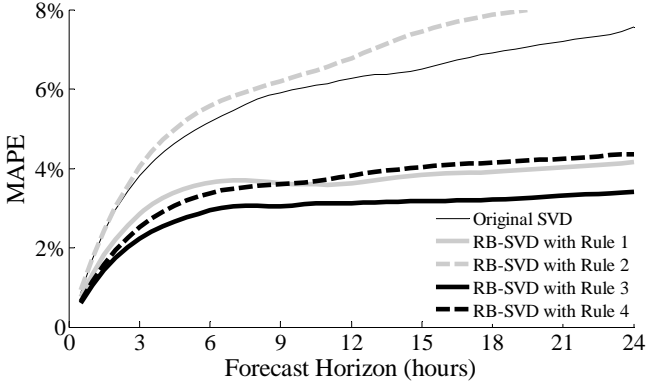

Fig. 8. MAPE across special days for the original and rule-based SVD.

We tested three further implementations of RB-ANN with Rule 1. (This rule was the most successful for RB-ANN). Each of these three further implementations included a set of new inputs, in addition to the lags described in Section III.C:
(1) RB-ANN with calendar data - We included as additional inputs: a counter running from 1 to 366 to indicate the day of the year; a counter running from 1 to 7 to indicate day of the week; a counter running from 1 to 336 to indicate half-hour of the week; and a counter running from 1 to 48 to indicate half-hour of day. Furthermore, following [14], we included sine and cosine function values of these 'counter variables'. We also included $I_{N_t}$ as defined previously in Section III.
(2) RB-ANN with extra lags – We used load at lags described in Section III.C, whereby $m_3(t)$ was chosen to be the same as the intrayear cycle length used in the original ANN method. We also used, as extra inputs, load at *LagA*, *LagB*, and *LagC* (as defined in Section III.C), whereby for only the special days, $m_3(t)$ was selected using a rule-based approach, while for the normal days, $m_3(t)$ was the same as the intrayear cycle length, as described above. We also included $I_{N_t}$ as an input.
(3) Separate ANN - Following [12], we built separate ANNs for normal and special days. The results for implementations

(1) and (2) were similar, with (3) being the worst. Hence, we only present below results for implementation (1).

In Fig. 9, for the special days, we plot the MAPE results for the best performing method from Figs. 5 to 8, along with RB-ANN with Rule 1 and calendar data, and the following three simple benchmarks: (i) seasonal random walk - load occurring on the same special day in the last year; (ii) seasonal moving average - mean of the load on the same special day in the last four years; and, (iii) recent Sunday - load on the most recent Sunday, this benchmark is used in [2]. The worst performing methods in Fig. 9 are the benchmarks.

In Fig. 9, of the four rule-based methods, the ANN results were the poorest, which is perhaps due to the absence of strong nonlinearity in the structure of the time series. Also, including calendar data as an additional input did not greatly improve the results of the RB-ANN method. The results for RB-HWT and RB-SARMA were similar with the latter slightly preferable. RB-SVD was competitive with RB-HWT, but less accurate than RB-SARMA. Given that HWT and SARMA are competitive, and are based on different modeling approaches, we consider combining forecasts from these two methods using a simple average. The figure shows that this led to greater accuracy than the individual models.

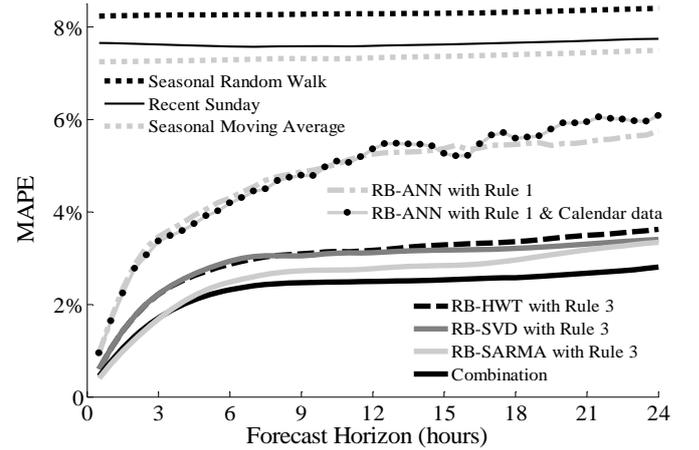

Fig. 9. MAPE across special days, for the best performing method from Fig. 5, 6, 7, and 8, along with the simple benchmarks.

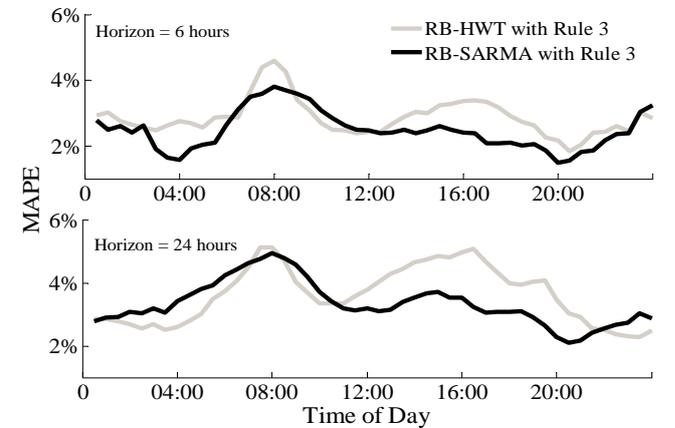

Fig. 10. MAPE across special days, using RB-HWT and RB-SARMA with Rule 3, plotted against different time of the day.

In Fig. 10, for six hour-ahead and one day-ahead prediction for just the special days, we plot the MAPE against the period of the day for RB-HWT and RB-SARMA, both with Rule 3. As expected, the large MAPE values correspond to periods of day when load changed by a relatively large amount.

For the rule-based HWT, SARMA and SVD methods, it is interesting to note that, for the special days, the MAPE values for Rule 3 are about half those of the corresponding original models that were not rule-based. Since anomalous observations constitute about five percent of the time series, the reported drop in error on special days can have significant practical impact. Also, modeling anomalous load not only results in lower errors on special days, but also aids modeling of load on normal days that lie in the vicinity of special days.

In Fig. 11, we report the forecasting performance of original and rule-based versions of SARMA and HWT across normal days. Although we developed the rule-based modifications with the aim of improving the forecasting of special days, we would hope that these modifications would not lead to poor forecast accuracy for the normal days. Therefore, in Fig. 11, it is reassuring to see that the results of the original and rule-based versions are similar for each method.

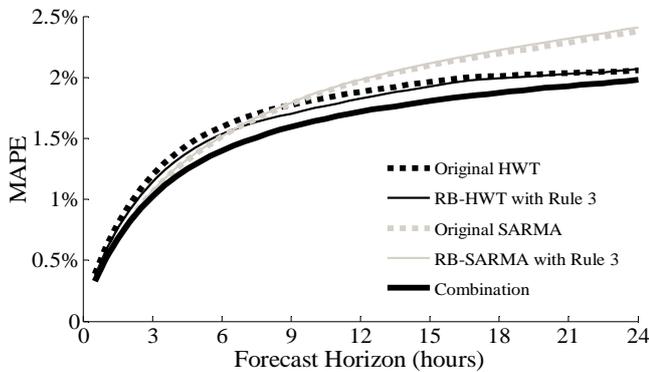

Fig. 11. MAPE across normal days, for RB-HWT and RB-SARMA with Rule 3, the corresponding original methods, and their combination.

## VI. SUMMARY AND CONCLUDING COMMENTS

In this paper, our aim has been to develop forecasting methods that can be used to effectively assist the central controller, and potentially be used for generating real-time online forecasts in an automated framework under both normal and anomalous load conditions. We showed that the accuracy of conventional methods can be improved for anomalous load forecasting by using a rule-based approach. We evaluated four different classes of univariate modeling approaches, when used in conjunction with four different rules. The new rule-based HWT method has the appeal of simplicity when compared with rule-based versions of SARMA, ANN and SVD methods, both in terms of the model structure and number of parameters. Overall, a combination of the rule-based HWT and SARMA methods was most accurate.

It would be worth investigating the performance of the rule-based methods, for different lengths of the time series. However, such a study would require a much longer time series compared to the one used in this work. A potentially interesting line of future work would be to set the annual lag differently for different times of the same special day. This would amount to using observations from different parts of different past special days to forecast a future special day.

## ACKNOWLEDGEMENT

The authors are grateful to the Editor and three anonymous referees for their useful comments and suggestions.

## REFERENCES

[1] D.W. Bunn, "Forecasting loads and prices in competitive power markets," *Proc. IEEE*, vol. 88, pp. 163-169, 2000.

[2] M. Smith, "Modeling and short-term forecasting of New South Wales electricity system load," *J. Bus. & Econ. Stat.*, vol. 18, pp. 465-478, 2000.

[3] J.W. Taylor, "Short-Term load forecasting with exponentially weighted methods," *IEEE Trans. Power Syst.*, vol. 27, pp. 458-464, 2012.

[4] S. Rahman, and R. Bhatnagar, "An expert system based algorithm for short term load forecast," *IEEE Trans. Power Syst.*, vol. 3, pp. 392-399, 1988.

[5] O. Hyde, and P.F. Hodnett, "Rule-based procedures in short-term electricity load forecasting," *IMA Jour. Math App. Buss. & Ind.*, vol. 5, pp. 131-141, 1993.

[6] O. Hyde, and P.F. Hodnett, "An adaptable automated procedure for short-term electricity load forecasting," *IEEE Trans. Power Syst.*, vol. 12, pp. 84-94, 1997.

[7] J. R. Cancelo, A. Espasa, and R. Grafe, "Forecasting the electricity load from one day to one week ahead for the Spanish system operator," *Int. J. Forecast.*, vol. 24, pp. 588-602, 2008.

[8] V. Dordonnat, S.J. Koopman, M. Ooms, A. Dessertaine, and J. Collet, "An hourly periodic state space model for modelling French national electricity load," *Int. J. Forecast.*, vol. 24, pp. 566-587, 2008.

[9] L.J. Soares, and M.C. Medeiros, "Modelling and forecasting short-term electricity load: A comparison of methods with an application to Brazilian data," *Int. J. Forecast.*, vol. 24, pp. 630-644, 2008.

[10] D. Srinivasan, C.S. Chang, and A.C. Liew, "Demand forecasting using fuzzy neural computation, with special emphasis on weekend and public holiday forecasting," *IEEE Trans. Power Syst.*, vol. 10, pp. 1897-1903, 1995.

[11] R. Lamedica, A. Prudenzi, M. Sforna, M. Caciotta, and V.O. Cencellli, "A neural network based technique for short-term forecasting of anomalous load periods," *IEEE Trans. Power Syst.*, vol. 11, pp. 1749-1756, 1996.

[12] K-H. Kim, H-S Youn, and Y-C. Kang, "Short-term load forecasting for special days in anomalous load conditions using neural networks and fuzzy inference method," *IEEE Trans. Power Syst.*, vol. 15, pp. 559-565, 2000.

[13] K-B. Song, Y-S. Baek, D.H. Hong, G. Jang, "Short-term load forecasting for the holidays using fuzzy linear regression method," *IEEE Trans. Power Syst.*, vol. 20, pp. 96-101, 2005.

[14] J.N. Fidalgo, and J.A. Peças Lopes, "Load forecasting performance enhancement when facing anomalous events," *IEEE Trans. Power Syst.*, vol. 20, pp. 408-415, 2005.

[15] D. Bunn, and G. Wright, "Interaction of judgmental and statistical forecasting methods: Issues and analysis," *Man. Sci.*, vol. 37, pp. 501-518, 1991.

[16] F. Collopy, and J.S. Armstrong, "Rule-based forecasting: development and validation of an expert systems approach to combining time series extrapolations," *Man. Sci.*, vol. 38, pp. 1394-1414, 1992.

[17] J.W. Taylor, "Triple seasonal methods for short-term electricity demand forecasting," *Eur. J. Oper. Res.*, vol. 204, pp. 139-152, 2010.

[18] A.F. Atiya, S.M. El-Shoura, S.I. Shaheen, and M.S. El-Sherif, "A comparison between neural-network forecasting techniques—case study: river flow forecasting," *IEEE Trans. Neural Networks*, vol. 10, pp. 402-409, 1999.

[19] H. Shen, and J.Z. Huang, "Interday forecasting and intraday updating of call center arrivals," *Manufacturing and Services Operations Management*, vol. 10, pp. 391-410, 2008.

**Siddharth Arora** is a doctoral student at the Saïd Business School, University of Oxford. His research interests include density forecasting and model combination with applications in energy.

**James W. Taylor** is a Professor of Decision Science at the Saïd Business School, University of Oxford. His research interests include energy forecasting, exponential smoothing and density forecasting.